\documentclass[aip, amsmath,amssymb,
 reprint,%
]{revtex4-1}
\usepackage[sort&compress]{natbib}
\usepackage{color}      
\usepackage{epsfig}
\newlength\figurewidth
\AtBeginDocument{\setlength\figurewidth{.5\linewidth}}
\let\vec\boldsymbol

\def\kT{\ensuremath{k_\text{B}T}}

\usepackage{glossaries}
\newacronym{mct}{MCT}{mode-coupling theory of the glass transition}
\newacronym{msd}{MSD}{mean-squared displacement}
\newacronym{ngp}{NGP}{non-Gaussian parameter}
\newacronym{isf}{ISF}{intermediate scattering function}
\newacronym{sisf}{SISF}{self-intermediate scattering function}
\newacronym{fse}{FSE}{finite-size effects}
\newacronym{dh}{DH}{dynamical heterogeneities}
\newacronym{vh}{vH}{van Hove}

\usepackage{todonotes}
\usepackage{comment}

\def\url#1{}

\begin{document}
\title{Using microrheology to study dynamical heterogeneities}
\date{\today}
\def\unial{\affiliation{Departamento de Qu\'{\i}mica y F\'\i{}sica, Universidad de Almer\'\i{}a, 04.120 Almer\'\i{}a, Spain}}
\def\unidd{\affiliation{Institut f\"ur Theoretische Physik, Heinrich-Heine-Universit\"at D\"usseldorf, 40225 D\"usseldorf, Germany}}
\def\dlrfm{\affiliation{Institute for Frontier Materials on Earth and in Space, German Aerospace Center (DLR), 51147 Cologne, Germany}}
\author{A.~M.~Puertas}\unial
\author{Thomas Voigtmann}\unidd\dlrfm

\begin{abstract}
Dynamical heterogeneities are one of the hallmarks of supercooled liquids, and their properties and relevance have been studied with theory, simulations and experiments. In this work, we propose to monitor the dynamics of tracer particles (passive microrheology) to analyze the dynamical heterogeneities in a system of hard colloids close to the glass transition density, using Langevin dynamics simulations and mode coupling theory. Different observables, typical in the study of the dynamical heterogeneities are adapted to be calculated from the trajectory of a single tracer particle. The tracer dynamics shows a transition from a regime where it is most decoupled from the bath for small tracer size to a strong coupling regime for large tracers. Both theory and simulations show that the non-Gaussian parameter of the tracer is maximal for tracer sizes at the crossover between both regimes, and is highly dependent on the bath density. The dynamic susceptibility is also studied, but this parameter shows a minor dependence on both the tracer size or the bath density. Finally, the existence of regions with different mobility is also studied with microrheology. Although the tracer trajectory indeed shows stages with increased mobility, the estimated size of the regions decreases with the bath density, contrary to the results from cluster analysis in the bulk. 
\end{abstract}

\maketitle

\section{Introduction}
\label{sect_intro}

\Gls{dh} are considered one of the hallmarks of the glass transition, adding to the dramatic increase of the viscosity and slowing down of the structural relaxation. These are caused by regions of the system with very different dynamics \cite{Ediger.2000}, and in some models of the glass transition, such as the early Adam-Gibbs scenario \cite{Adam.1965} or the random first order transition theory \cite{Kirkpatrick.1989}, they are the precursors of the glass transition. Indeed, \gls{dh} and regions of different mobility have been identified in experiments of many different systems close to the glass transition, both in simple or complex molecular systems \cite{Patterson.1994,Dasini.2026}, or in colloids \cite{Weeks.2000,Narumi.2011}. However, these regions are microscopic, and their observation in the bulk requires specific techniques, such as time
resolved correlation \cite{Cipelletti.2003, Duri.2005, Kanayama.2022}. 

Microrheology, on the other hand probes the mechanical behaviour of a system at the microscopic scale. Microrheology was proposed around 30 years ago to study the viscoelastic behaviour of soft matter samples \cite{rheology:Mason.1995,Cicuta.2007,FurstSquires}. There, a colloidal tracer is introduced in the sample and its dynamics is monitored; in active microrheology, the tracer is pulled by an external force, whereas in passive microrheology it diffuses solely subjected to thermal and density fluctuations of the bath. Although the method was originally proposed to infer the viscoelastic properties of the bath, it was soon realised that the interpretation was not straightforward \cite{Squires.2005,Squires.2008,Gazuz.2009,Puertas.2014}. But its possibilities were also larger than expected, as it offered the opportunity to study the mechanical properties of other complex systems, such as living cells \cite{Wirtz.2009}, or glass and gel aging \cite{Rich.2011}.

Dynamical heterogeneities at the microscopic level have been determined and quantified mostly in simulations. Initially, the deviations from the Gaussian displacement distribution (van Hove function), were measured by the non-Gaussian parameter \cite{Kob.1997,Donati.1999}. Likewise, a dynamical susceptibility, as a four point correlation function, was proposed, revealing that heterogeneities are maximal at the timescale of the structural relaxation \cite{Glotzer.2000,Lacevic.2003,glass-simulation:Coniglio.2008}. The existence of clusters of particles with different mobility has been also tested \cite{Hedges.2009,Tahaei.2023}, but these remain microscopic. Microrheology has been also used to
study the heterogeneities \cite{Valentine.2001,Sarangapani.2008,Nakamura.2025}. This not only offers the possibility to test heterogeneities at the microscopic level, but also to modify the probe length scale in
soft matter systems, where hierarchical structures are an intrinsic characteristic \cite{Sarangapani.2008,Nakamura.2025}, and also in atomic systems \cite{Paeng.2015}. Interestingly, in polymer melts, experiments with tracers of different sizes have shown that rheological properties are reproduced with large tracers, whereas the small tracers are more useful to study the heterogeneities \cite{Valentine.2001,rheology:Lu.2002,Grabowski.2014}. However, a systematic study of a system with well characterized \gls{dh} with tracers of different sizes has not been performed.

In this paper we aim close this gap by using passive microrheology to study the \gls{dh} of a system of quasi-hard spheres using simulations. Single tracers have been used with sizes, $a_t$, ranging from $a_t/a = 0.1667$ to $a_t/a = 4$, where $a$ is the mean bath particle radius. The system's non-Gaussian parameter and susceptibility are analyzed, as well as a measure of the size of the regions with different dynamics. Then, these observables are calculated from the tracer trajectory exchanging ensemble average for time-origin average. The results show that the most prominent \gls{dh} in the tracer are obtained for sizes smaller than the bath particles radius, with a maximum shifting its position with bath density. We also contrast these simulation data with theoretical results from the \gls{mct}, a theory that focuses on microscopic \gls{dh} even though it does not fully predict the macroscopic ones correctly. Furthermore, the regions with increased mobility are estimated identifying portions of the tracer trajectory with larger displacement than average. The size of the regions determined in this way provide a lower bound for their true magnitude, but clearly probe their presence.

\section{Simulation details}
\label{sect_sim}

We simulate a colloidal system in a cubic box with periodic boundary conditions composed by a bath with $N-1$ quasi-hard particles, and one tracer particle (labeled with $j=1$ hereafter). This system has been described previously (see e.g. Ref.~\cite{Orts.2023}).

The micoscopic dynamics of all particles follows the Langevin equation \cite{Dhont.1996}, which reads, for particle $j$:

\begin{equation}
m_j \frac{d^2\, {\bf r}_j}{dt^2}\:=\:  - \gamma_j \frac{d\, {\bf r}_j}{dt} + {\bf f}_j(t) + \sum_{i\neq j} {\bf F}_{ij}
\label{Langevin}
\end{equation}

\noindent where $m_j$ is the particle mass; $\gamma_j$ is the friction coefficient with the solvent: $\gamma_j=\gamma_0 a_j$ with $\gamma_0$ constant for all particles, and $a_j$ the particle radius; ${\bf f}_j$ is a random Brownian force which fulfills the fluctuation-dissipation theorem, $\langle {\bf f}_j(t) \cdot {\bf f}_j(t') \rangle = 6 \kT \gamma_j \delta(t-t')$, where $\kT$ is the thermal energy and $\delta(x)$ is the Dirac-delta symbol \cite{Dhont.1996}; and ${\bf F}_{ij}$ is the interaction force between particles $i$ and $j$. The latter derives from the central inverse-power potential:

\begin{equation}
V({\bf r})\:=\:\kT \left( \frac{r}{a_{ij}} \right)^{-36} \label{potential}
\end{equation}

\noindent with $r=\left| {\bf r} \right|$ the center to center distance between the particles $i$ and $j$, and $a_{ij}=a_i+a_j$. This potential is steep enough to mimic effectively the dynamics of the hard sphere system \cite{Lange.2009}. A slightly polydisperse bath is simulated to avoid crystallization at high density; the particle radii are drawn from a flat distribution of width $0.2a$, with $a$ the mean radius of the bath particles, but all particles have the same mass $m$. In the simulations, the mean bath particle radius, $a$, mass particle $m$, and thermal energy $\kT$ are the units for length, mass and energy, respectively. The equations of motion have been integrated following the Heun algorithm \cite{Paul.1995}, with a time step of $\delta t=5\cdot 10^{-4} \,a \sqrt{m/\kT}$, and for the friction force, we set $\gamma_0=5\,\sqrt{m\kT}/a$. The tracer has the same mass as the bath particles, and its radius has been varied from $\delta=a_t/a=0.1667$ to $\delta=a_t/a=4$, given in terms of the ratio of the tracer particle radius to the average bath particle radius. The volume fraction of the system approaches the glass transition, ranging up to $0.57$, where the \gls{dh} are expected to be relevant. Previous analysis of the diffusion coefficient and structural relaxation, in comparison with MCT, determined the glass transition of the bath at $\phi_c=0.596$ \cite{Voigtmann.2004b,Weysser.2010}.

\Gls{fse} have been previously studied in this system \cite{Orts.2023}. Using the tracer diffusion coefficient, \gls{fse} were found for tracers larger than the bath particles, i.e. $\delta>1$, and saturate for system sizes above $N\approx 8000$ particles. Further analysis confirmed these results for the non-Gaussian parameter and dynamical susceptibility. Therefore, in this case systems with $N=1000$ particles were considered for tracers with $\delta \leq 1$, and $N=8000$ particles for larger tracers. 

We use several quantities to analyze the dynamics of the bath and tracer, and study \gls{dh}. The \gls{ngp} is the simplest quantity, as it measures the deviation of the self part of the \gls{vh} function from Gaussian behaviour. For this purpose, the ratio of the fourth to second moments is used \cite{Rahman.1964}:
\begin{equation}
\alpha_2(\tau)\:=\: \frac{3}{5} \frac{\langle \delta r^4(\tau) \rangle}{\langle \delta r^2(\tau)\rangle^2} - 1 \label{eq:alpha2-tracer}
\end{equation}
where $\langle \cdots \rangle$ refers to the ensemble average (average over all particles in the system and time origin in the simulations). This parameter tends to zero for short and long times, where short- and long-time diffusion are observed, respectively, and the \gls{vh} function recovers its Gaussian shape. In microrheology, only the tracer is observed, and $\alpha_2$ is calculated only with its dynamics. Therefore, the average is restricted only to time-origin averages, as a single particle is monitored.

A more global parameter was introduced by Glotzer et al. \cite{Glotzer.2000,Lacevic.2003}, as a dynamical susceptibility, which measures the variability of the relaxation dynamics. In this case, the variance of the density autocorrelation function is used:
\begin{equation}
\chi_4(\tau) = \left\langle \left[\hat\Phi_q^s(\tau)\right]^2 \right\rangle - \left[\left\langle \hat\Phi_q^s(\tau)\right\rangle \right]^2
\end{equation}
where $\hat\Phi_q^s(\tau)$ is a density-fluctuation pair,
$\hat\Phi_q^s (\tau) = \frac{1}{N} \sum_i \cos \left[ \vec{q} \cdot (\vec{r}_i(\tau) - \vec{r}_i(0)) \right]$ for wavevector $\vec{q}$,
such that the ensemble average $\langle\hat\Phi_q^s(\tau)\rangle=\Phi_q^s(\tau)$ yields the
self-part of the \gls{isf}, \gls{sisf}.
The dynamical susceptibility starts from zero at short times, and typically shows a maximum at the relaxation time, decaying to zero or a finite value depending on the details of the particle interactions \cite{glass-simulation:Coniglio.2008}.

For the tracer, the scattering function is defined as $\hat\Phi_q^t (\tau) = \cos \left[ \vec{q} \cdot (\vec{r}_t(\tau) - \vec{r}_t(0)) \right]$, 
for a single trajectory, and the susceptibility is:
\begin{equation}
\chi_4^t(\tau) = \left\langle \left[\hat\Phi_q^t(\tau)\right]^2 \right\rangle - \left\langle \hat\Phi_q^t(\tau) \right\rangle^2 \label{eq-chi4-tracer}
\end{equation}
requiring averaging over different trajectories and/or time origins.

Finally, to identify regions with different mobility, particles with a displacement in a given interval, $\Delta t_*$, well above the average (fast particles) are sought; here we take $\Delta t_*$ equal to the time when the NGP is maximal, and particle $j$ is fast if $\Delta r_j^2(\Delta t_*) > 4 \langle \Delta r^2(\Delta t_*)\rangle$. Two fast particles belong to the same cluster if their separation is smaller than a threshold distance (here we take twice the localization length). The size distribution of clusters of fast particles in undercooled fluids show typically a maximum for isolated  particles, but the distribution extends to larger sizes as the density increases. In microrheology, the tracer trajectory is divided in pieces of duration $\Delta t_*$ and it is considered that the tracer is in a fast region if its displacement in $\Delta t_*$ is larger than twice the tracer mean tracer displacement in the same time interval. The distance travelled within the fast region gives a first estimate of its size. However, since these fast regions are very heterogeneous \cite{glass-confined:Peng.2022}, this method is expected to underestimate the true size of the fast regions.

\section{Theory Details}
\label{sec:theory}

The \gls{mct} predicts the slow structural relaxation of density fluctuations
close to the glass transition, based on the static structure factors.
It allows to also predict the tracer dynamics, and from that the
\gls{msd} and the \gls{ngp}. We compare the predictions of the theory
to the simulation qualitatively, focusing on the dependence on the
tracer size.

The theory for the bulk has been established in great detail before
\cite{Goetze.2009,Janssen.2018}, therefore we
only give a brief summary of the salient details.
We compare the simulations to overdamped Brownian dynamics of (monodisperse)
hard spheres as a good model system for the simulation. We will point
to discrepancies that likely stem from the simulation system being
polydisperse, where appropriate.
The Mori-Zwanzig evolution equation of the normalized collective density
correlation function $\phi_q(t)=\Phi_q(t)/S_q$ reads
\begin{equation}
\tau_0(q)\partial_t\phi_q(t)+\phi_q(t)
  +\int_0^tm_q(t-t')\dot\phi_q(t')\,dt'=0\,.
\end{equation}
Here $q=|\vec q|$, and $\tau_0(q)=S_q/q^2D_0$ sets the time scale through the
short-time diffusion coefficient $D_0$.
The memory kernel is approximated in \gls{mct} as a functional of the
correlation functions, and one obtains after a standard set of
approximations
\begin{equation}
  m_q(t)=\frac{n}{16\pi^3}\frac1{q^2}\int d\vec k\,S_qS_kS_p
  V_{qkp}
  \phi_k(t)\phi_p(t)
\end{equation}
with $\vec p=\vec q-\vec k$ and
$V_{qkp}=[(\vec q\cdot\vec k)c_k+(\vec q\cdot\vec p)c_p]^2$.
The direct correlation function
$c_k$ is given by the static structure factor of the system, $S_k$, that
consequently becomes the sole input to the theory besides the number
density $n$.

The theory has been also extended to the case of passive and active microrheology \cite{Gazuz.2009,Puertas.2014}. In the former case, the \gls{mct} equation for the tracer particle mimics those for the bath:
\begin{equation}
\tau^s_0(q)\partial_t\phi^s_q(t)+\phi^s_q(t)
  +\int_0^tm^s_q(t-t')\dot\phi^s_q(t')\,dt'=0\,,
\end{equation}
with $\tau^s_0(q)=1/q^2D_0^s$, and the memory kernel
\begin{equation}
  m^s_q(t)=\frac{n}{(2\pi)^3}\frac1{q^2}\int d\vec k\,S_k
  \left[(\vec q\cdot\vec k)c^s_k\right]^2\phi_k(t)\phi^s_p(t)
\end{equation}
with the tracer-particle direct correlation function $c^s_k$. In the case of a tracer that is identical to the bath particles,
$c^s_k=c_k$; else $c^s_k$ encodes the size difference.

From the \gls{sisf} one obtains the \gls{msd} and the \gls{ngp} in the
low-$q$ limit. For numerical reasons it is advisable to perform this limit
analytically; one obtains \cite{Fuchs.1998}
\begin{equation}
  \delta r^2(t)+D_0^s\int_0^tm^{(0)}(t-t')\delta r^2(t')\,dt'=6D_0^st
\end{equation}
with the memory kernel
$m^{(0)}(t)=\lim_{q\to0}q^2m^s_q(t)$,
\begin{equation}
  m^{(0)}(t) 
  =\frac{n}{6\pi^2}\int_0^\infty dk\,S_k\left[k^2c^s_k\right]^2
  \phi_k(t)\phi^s_k(t)\,.
\end{equation}
The \gls{ngp} $\alpha_2(t)$ follows from the next-to-leading term in low $q$
and obeys \cite{Fuchs.1998,Mayr.1998},
\begin{multline}
  \left[1+\alpha_2(t)\right]\delta r^2(t)^2
  \\
  +D_0^s\int_0^tm^{(0)}(t-t')\left[1+\alpha_2(t')\right]\delta r^2(t')^2\,dt'
  \\ =6D_0^s\int_0^t\left[2+m^{(2)}(t-t')\right]\delta r^2(t')\,dt'
\end{multline}
with a further kernel
\begin{multline}
  m^{(2)}(t)=\frac{n}{10\pi^2}\int dk\,S_k\left[k^2c^s_k\right]^2
    \phi_k(t)\times\\ \times\left[\frac{\partial^2\phi^s_k(t)}{\partial k^2}
    +\frac2{3k}\frac{\partial\phi^s_k(t)}{\partial k}\right]
    \,.
\end{multline}

The equations are solved using a straightforward integration scheme,
using a time-domain discretization that consists of a staggered grid
of equidistant points, with a repeated doubling of the step size in order
to obtain solutions of large time spans. The results in this paper
have been obtained with the open-source code \textsc{mctpy}
\cite{mctpy}, discretizing wave numbers onto a grid
$q_ia=0.1+0.2i$ with $100$ points (implying a cutoff at
$qa=19.9$).
The static-structure quantities have been evaluated within the
Percus-Yevick approximation from which one obtains analytical expressions
for $c_k(n)$ and $c_k^s(n,\delta)$ \cite{Baxter.1970,Wertheim.1963}. Note that \gls{mct} predicts the glass transition for this system at a volume fraction $\phi_c\approx 0.516$ \cite{Franosch.1997}, i.e., at a lower volume fraction than in the simulation. We follow the established practice of comparing volume fractions roughly according to their respective distance to the glass-transition point.

\section{Results and discussions}
\label{results}

We present first the analysis of the \gls{dh} of the bath with different densities, as studied typically with the \gls{ngp}, the susceptibility calculated as the standard deviation of the \gls{sisf}, and the cluster analysis of fast particles. Then, the dynamics of the tracer is analysed using the same observables. The comparison between both methods allows us to determine how microrheology can be used to probe \gls{dh}.

\subsection{Dynamic heterogeneities of the bath}

\begin{figure}
\psfig{file=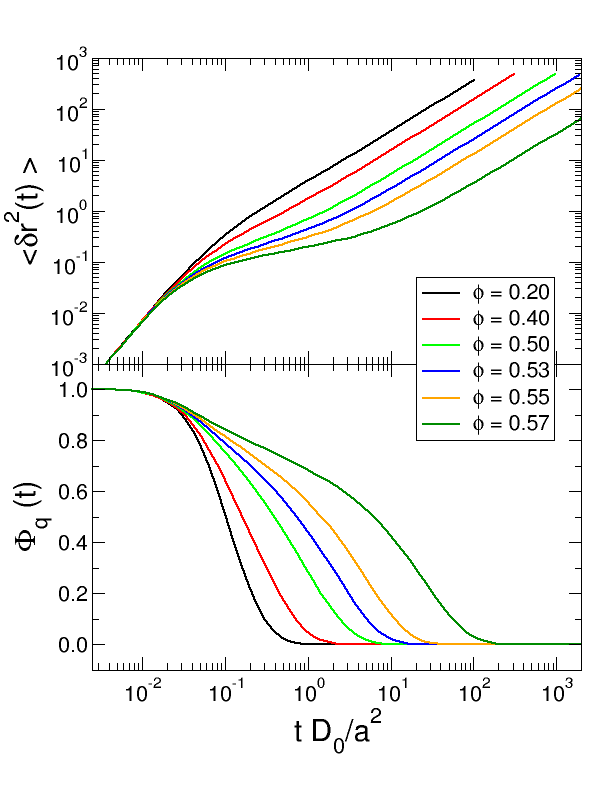,width=0.95\figurewidth}
\caption{Self part of the intermediate scattering function, $\phi_q(t)$, (top panel) and mean squared displacement (bottom panel) of the bath for different volume fractions, as labeled. \label{fsqt-msd-bath}}
\end{figure}

As a reference, Fig.~\ref{fsqt-msd-bath} shows the \gls{msd} and the \gls{sisf} for different packing fructions approaching the glass transition; for the \gls{sisf} we chose a wave vector corresponding to $qa=3.5$, close to the first-neighbor peak of the static structure factor. The developmend of the intermediate plateau in both quantities, as well as the stretched decay of the \gls{sisf} are clearly identifiable.

\begin{figure}
\psfig{file=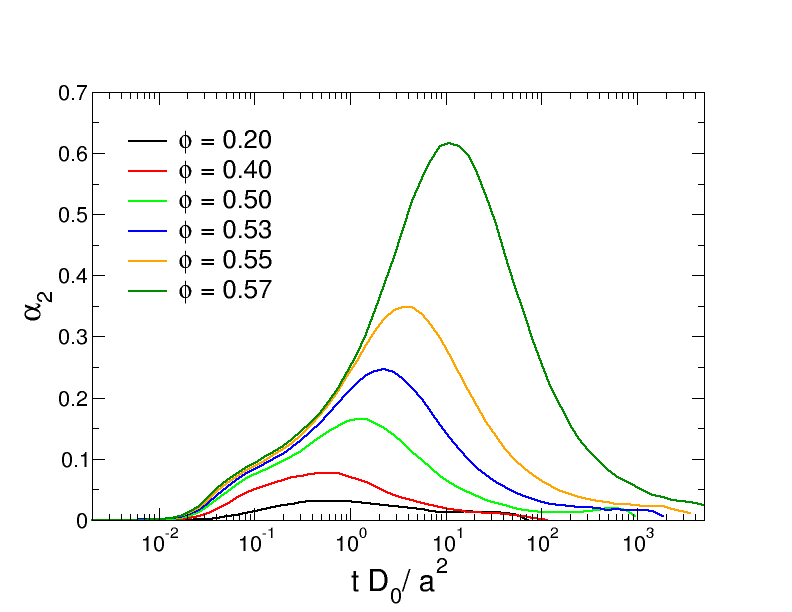,width=0.95\figurewidth}\\
\psfig{file=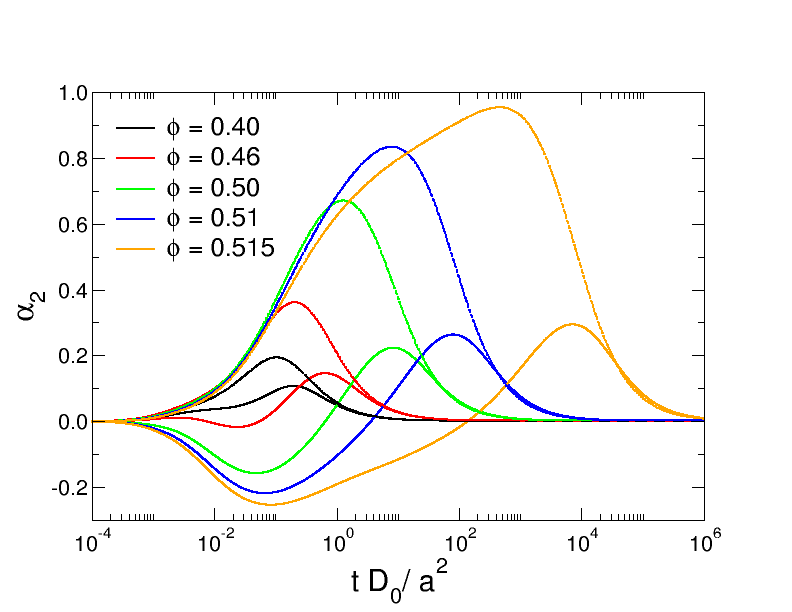,width=0.95\figurewidth}
\caption{Non-Gaussian parameter of the bath for different volume fractions, as labeled. Top: simulation, bottom: MCT where solid lines are $\delta=1$, and dashed lines $\delta=0.6$ \label{alpha2-bath}}
\end{figure}

The \gls{ngp} is presented in Fig. \ref{alpha2-bath} for the same volume fractions as shown in Fig. \ref{fsqt-msd-bath} (simulations in the upper panel, theory in the lower one). As expected, the parameter shows a peak in the time scale where structural relaxation occurs (the correlation function decays from the plateau, or the \gls{msd} leaves the plateau of the localization length). This peak moves to longer times and, more importantly, grows with the volume fraction, indicating that the particle displacements during structural relaxation deviate from the Gaussian distribution more strongly. 

The simulation results broadly follow what is known from previous simulations, such as Lennard-Jones particles \cite{Kob.1995,Kob.1997} or hard spheres \cite{Doliwa.1999, Yamaguchi.2001}. Note that in the Kob-Andersen mixture, the \gls{ngp} appears slightly more pronounced \cite{Kob.1995}. This shows that the \gls{ngp} is a highly sensitive parameter, influenced by the details of the particle distribution (a binary mixture in the Kob-Andersen system, compared to a continuous polydispersity distribution studied here).

Figure~\ref{alpha2-bath} also shows the results from \gls{mct} (bottom panel). These calculations repeat those of \Citeauthor{Fuchs.1998} \cite{Fuchs.1998}. Comparing to the simulation one notes that the main feature of a strong peak in the structural-relaxation regime is captured by \gls{mct}, as is the growth of that peak as one approaches the glass transition. Note that simulations often see a continued increase in the regime of the ideal glass of \gls{mct}, cf.\ Ref.~\cite{Laudicina.2024}, while \gls{mct} predicts the peak height in the \gls{ngp} to approach an asymptotic constant at its glass transition.

In the \gls{mct} results for $\delta=1$ (continuous lines) one also notes a strong negative dip corresponding to times of the early plateau regime, which is not seen in the simulations. This appears to be a systematic defect of \gls{mct}, as has been analyzed in the high-dimensional limit \cite{Laudicina.2024}. In fact, the asymptotic plateau value of the \gls{ngp} within \gls{mct} for the monodisperse hard-sphere system is negative \cite{Fuchs.1998}. A negative \gls{ngp} has been attributed to the cage effects: nearest-neighbor cages should suppress large displacements, and such confinement gives negative \gls{ngp}. The absence of this regime in the \gls{ngp} obtained from simulation could stem from the over-estimation of the cage effect by \gls{mct}, a known deficiency of the theory.
However, this is strongly tracer-size dependent: The dashed lines in
Fig.~\ref{alpha2-bath} show \gls{mct} results for a smaller tracer, $\delta=0.6$;
here, the region of negative values has completely disappeared. Also, the
peak in the \gls{ngp} is much stronger, and quantitatively closer to that
found in the simulation, as is the common envelope of the curves that emerges
as the packing fraction is increased. The case of $\delta\neq 1$ is further discussed in the next section.

\begin{figure}
\psfig{file=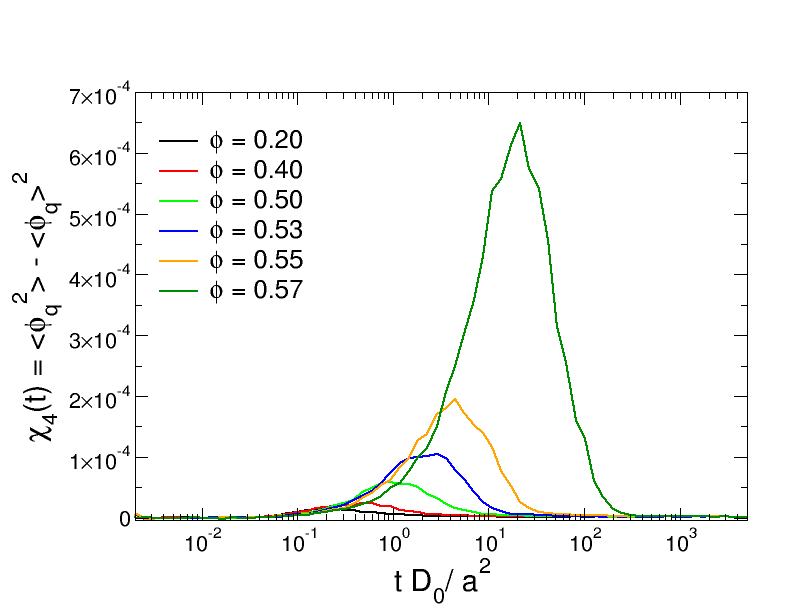,width=0.95\figurewidth}
\caption{Susceptibility of the bath for different volume fractions, as labeled. \label{chi4-bath}}
\end{figure}

A similar picture can be drawn studying the susceptibility from the intermediate scattering function, with a peak in the time scale of the structural relaxation. Fig. \ref{chi4-bath} shows the simulation results for our system, with an intense maximum for the largest volume fraction, and decaying to zero for long times. These results are in agreement with similar systems, such as hard spheres or Lennard Jones particles or systems with moderately strong short-range attraction \cite{Glotzer.2000, Reichman.2005, Charbonneau.2007,Brambilla.2009}.

In addition to these global quantities characterizing the heterogenous dynamics, a particle-scale analysis can be performed in the simulation by identifying clusters of fast particles. As mentioned above, the latter are defined as particles with a squared displacement larger than four times the \gls{msd} for the time where $\alpha_2$ is maximum, $\Delta t^*$, and two fast particles are considered neighbors when their surface-to-surface distance is lower than $0.2a$, twice the localization length. 

\begin{figure}
\psfig{file=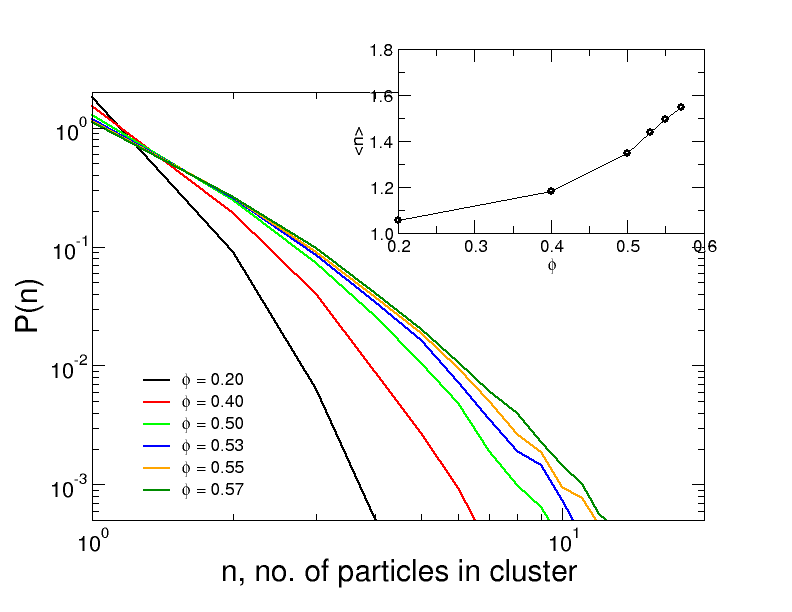,width=0.95\figurewidth}
\psfig{file=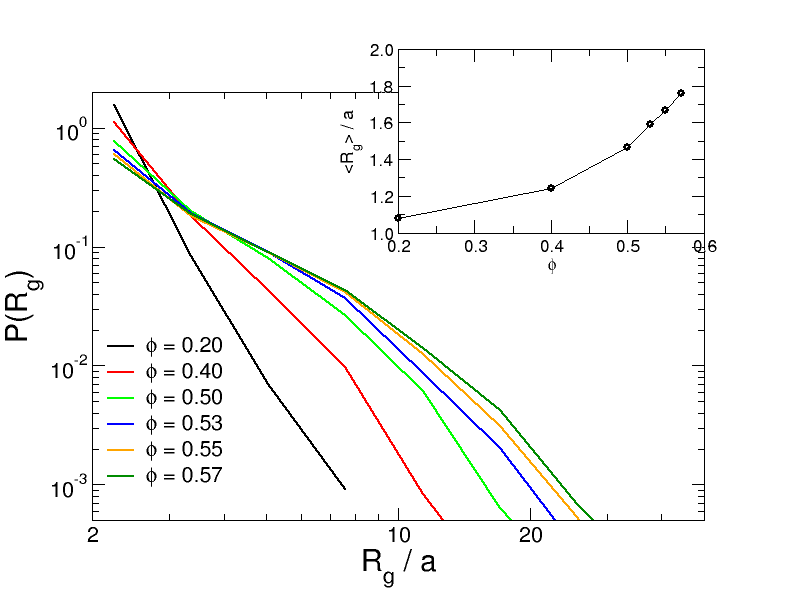,width=0.95\figurewidth}
\caption{Distribution of cluster sizes of fast particles as a funtion of the number of particles (upper panel) and radius of gyration (lower panel). \label{cluster-distrib-bulk}}
\end{figure}

The cluster size distributions of these clusters of fast particles are shown in the upper panel of Fig. \ref{cluster-distrib-bulk} for different volume fractions; single particle clusters are the most common type for all densities, but the distribution extends to larger clusters for higher density. The average number of particles per cluster (shown in the inset grows continuously up to $\langle n \rangle \sim 1.6$ for $\phi=0.57$). The cluster size distribution in terms of the radius of gyration is shown in the lower panel. Again, small clusters (single particles) are dominant, and the distribution extends to larger sizes for higher density.

\begin{figure*}
\psfig{file=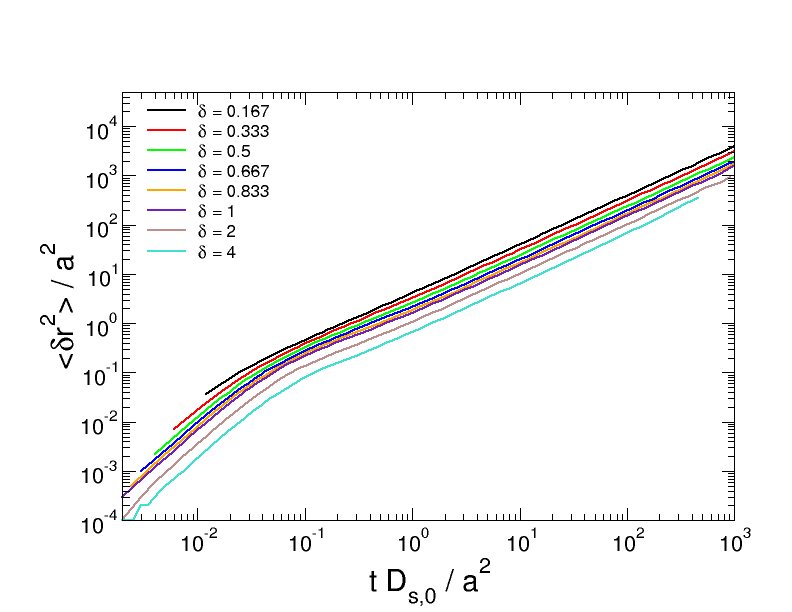,width=0.9\figurewidth}
\psfig{file=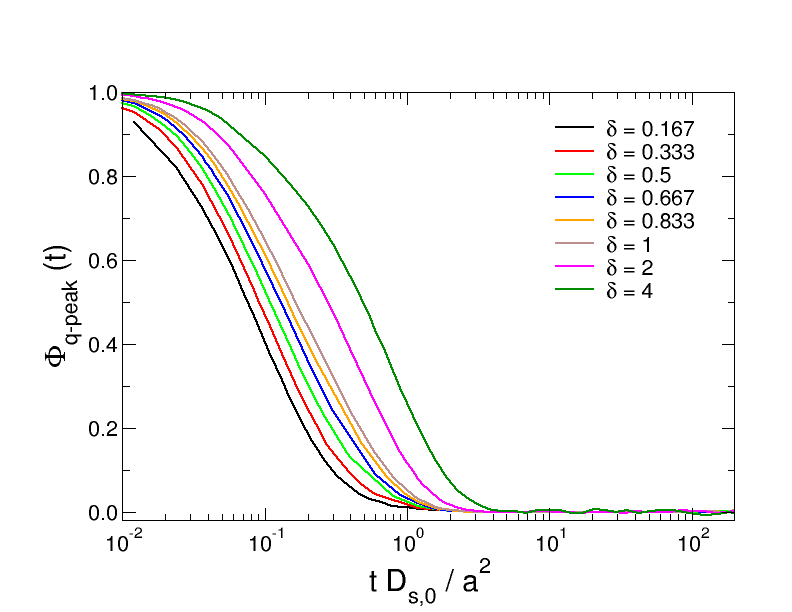,width=0.9\figurewidth}
\psfig{file=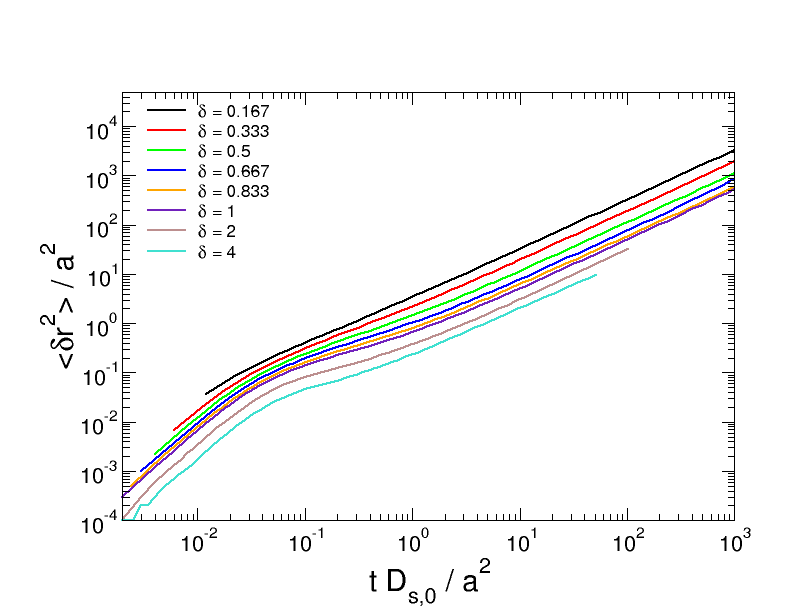,width=0.9\figurewidth}
\psfig{file=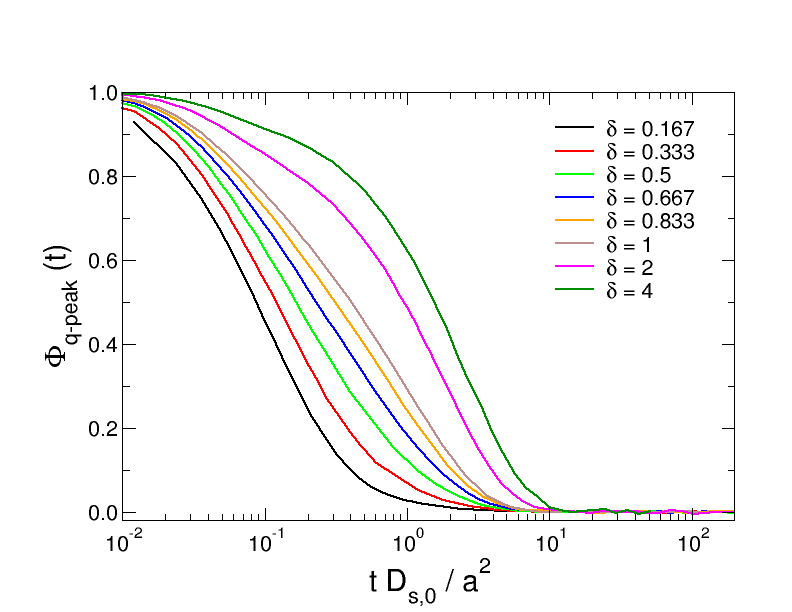,width=0.9\figurewidth}
\psfig{file=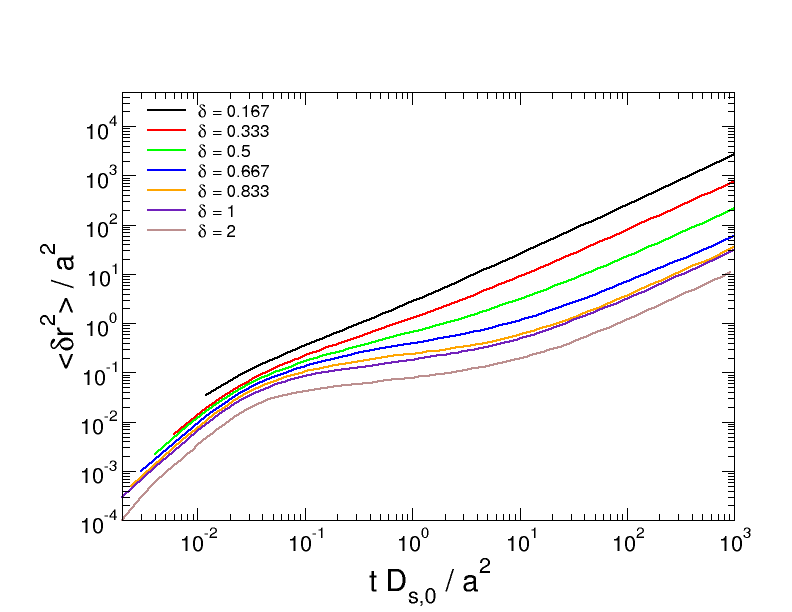,width=0.9\figurewidth}
\psfig{file=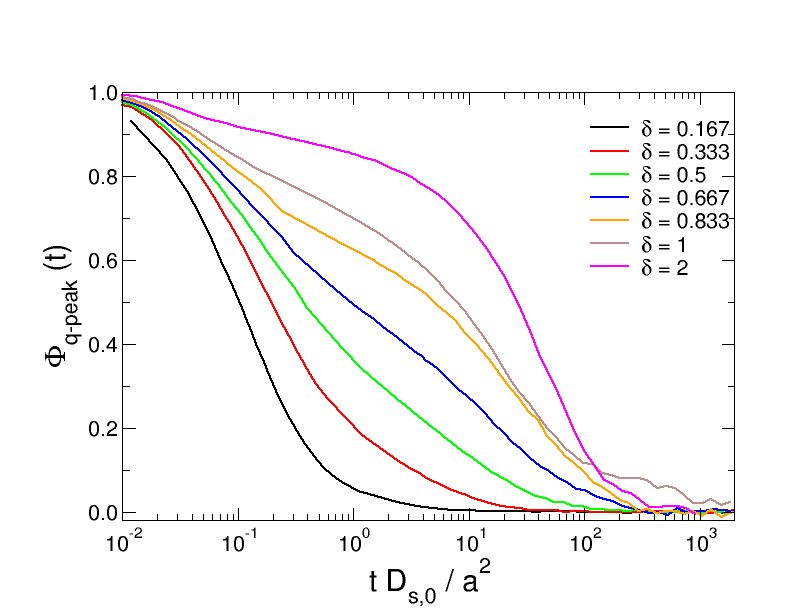,width=0.9\figurewidth}
\caption{Tracer mean squared displacement (left panels) and intermediate scattering function (panels in the right) for different bath density; $\phi=0.40$ in the upper panels, $\phi=0.50$ in the intermediate ones and $\phi=0.57$ in the bottom. \label{MSD-tracer}}
\end{figure*}

\subsection{Dynamic heterogeneities in the tracer dynamics}

We start studying the tracer \gls{msd} and intermediate scattering function, $\Phi_q^t(\tau)$, to aid a better interpretation of the parameters characterizing the heterogenous dynamics; different bath densities are shown in the panels of Fig. \ref{MSD-tracer}. Because big tracers have larger friction coefficient with the solvent, the single particle diffusion coefficient varies with the tracer size, causing the spread of the curves for the \gls{msd} at short times.


As a general trend, small tracers diffuse faster, and correspondingly show faster-decaying \gls{sisf}. The effect becomes more and more pronounced for densities closer to the glass transition. Fig.~\ref{MSD-tracer} (bottom right) shows that in the vicinity of dynamical arrest, changing the tracer size $a_t$ has a qualitative effect: while large tracers follow the familiar two-step relaxation pattern of the bath particles, sufficiently small tracers decouple dynamically, showing no indication of the intermediate-time plateau. As a side comment, we note that rattling of the large tracers in the cage of neighbors is smeared out by the large friction with the solvent \cite{Orts.2023}.

The decoupling scenario is in principle predicted by \gls{mct}
\cite{Bosse.1987,Bosse.1995,glass-theory:Voigtmann.2011,Sentjabrskaja.2016}.
There are caveats due to the fact that the way the standard \gls{mct}
for glass transitions deals with delocalization transitions that involve
a divergent length scale, is flawed \cite{Schnyder.2011}. However,
the qualitative scenario is robust in the sense that the small tracers
couple less and less to the standard two-step relaxation scenario of the
bulk as the size ratio is decreased.

\begin{figure}
\includegraphics[width=.8\figurewidth]{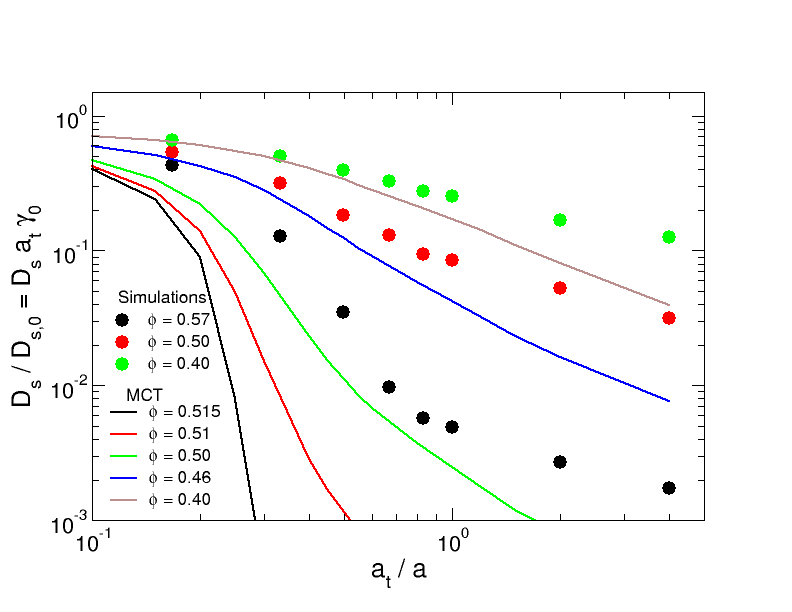}\\
\includegraphics[width=.8\figurewidth]{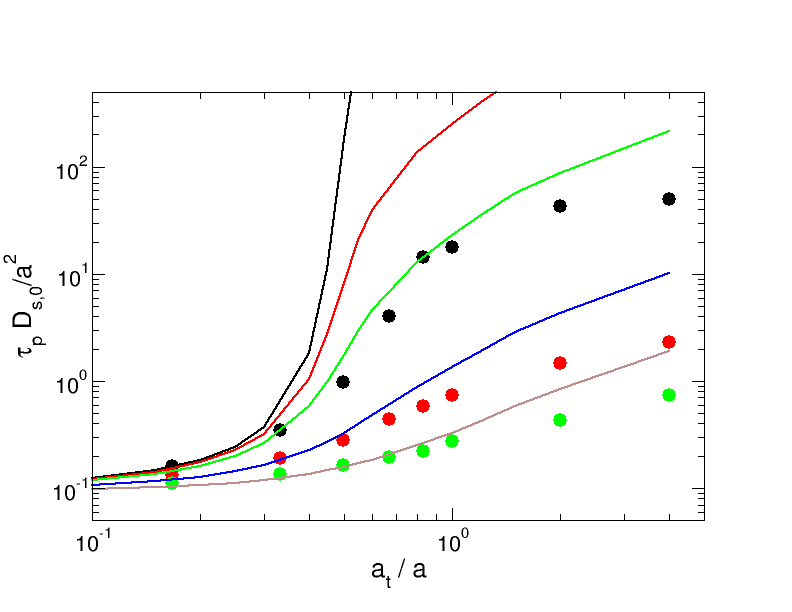}
\caption{Tracer diffusion coefficient obtained from the long time slope of the MSD and relaxation times from the intermediate scattering function for different bath volume fractions, as labeled. Symbols are simulation data, while linest represent MCT results.\label{diff-tracer}}
\end{figure}

The dynamical decoupling is exemplified by the tracers' long-time diffusion coefficients, shown in the upper panel of Fig.~\ref{diff-tracer} as a function of tracer size
for different volume fractions. Note that especially for the largest volume fraction studied in the simulation, $\phi=0.57$, the diffusion coefficient shows a sharp decrease around $\delta\sim 0.5$; at lower volume fractions, this sharp decrease is smeared out.
The same conclusions can be drawn from the relaxation times of the \gls{sisf}, obtained as the times where the function decays to the value $1/e$ (bottom
panel of Fig.~\ref{diff-tracer}).

The decoupling can be interpreted in the framework of \gls{mct}.
As shown in Fig.~\ref{diff-tracer}, also the theory reproduces this sharp
change. Within the theory, as the packing fraction is increased, one
approaches the situation of a (almost) frozen matrix in which  particles
are only arrested if they are sufficiently large. Hence, the sharp
decrease in the $D_L$ is the precursor of a step function from finite values
for $\delta<\delta_c$ (particles below a certain size $\delta_c$ remain
mobile even in the ideal glass), to zero for $\delta>\delta_c$.
Within \gls{mct}, we obtain $\delta_c\approx0.15$ \cite{Bosse.1987}.
The precise value of $\delta_c$ in the simulation data is hard to determine,
since the ideal glass transition is an avoided transition, and also due to the
size polydispersity.
In summary, the drop indicates a change in the diffusion mechanism from
fast diffusion of small tracers in an almost frozen matrix, to structural
relaxation of the bath.
The structural relaxation time of the \gls{sisf} (lower panel
of Fig.~\ref{diff-tracer}) shows the same qualitative behavior: a sharp
step evolves around $\delta_c$, as the packing fraction is increased.
Again, a shift in the determination of $\delta_c$ becomes apparent in
comparing simulation and theory.

Note that the limit of large $\delta$ should follow the Stokes-Einstein
scaling, $D\sim1/a_t$, following the direct proportionality between the drag coefficient and particle size. This is approached in the simulations (cf.\ the
dashed line representing the asymptotic behavior in Fig.~\ref{diff-tracer}).
In the theory, this asymptote is not correctly reproduced, since it misses
contributions from the transversal flow \cite{Orts.2023}. This flaw of
the theory notwithstanding, the agreement for the localization-induced
behavior at small $\delta$ is remarkably good.

\begin{figure}
\includegraphics[width=0.45\textwidth]{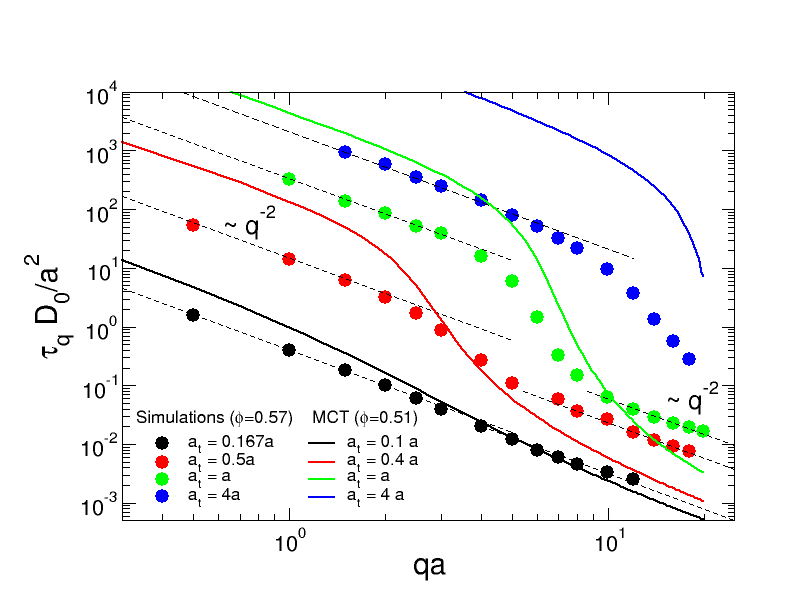}
\caption{Time scale of the relaxation of the tracer scattering function as a function of the wavevector modulus, $q$, for a bath of $\phi=0.57$. Simulations and theory are shown, as labeled. \label{tauq-tracer}}
\end{figure}

The tracer position correlation function, $\hat{\Phi}_q^t(\tau)$, can be also used to monitor its dynamics. The time scale for its decay, i.e. the time $\tau_q$ when $\hat{\Phi}_q^t(\tau_q)=1/e$, as a function of the wavevector modulus $q$ is presented in Fig. \ref{tauq-tracer} for $\phi=0.57$ and different tracer sizes; simulation data are indicated by the points and theory by the lines. In all cases, diffusion is attained for large length scales (small-$q$) as well as small distances (large-$q$), as noted by the $q^{-2}$ behaviour, indicated by the dashed lines. These two diffusive regimes are separated by a $q$-region where $\tau_q$ decreases much faster as a function of $q$. Notably, for small tracers, $\delta=0.167$, the difference between the diffusive regimes is very small, indicating that the diffusion of a small tracer is unaffected by the length scales of the bath particles. Instead, it sees a scale-free porous structure through which it diffuses \cite{Sentjabrskaja.2016},
in agreement with the interpretation from Fig. \ref{diff-tracer}. In contrast, for large tracers the separation between microscopic and long-distance diffusive regimes increases notably, as the effect of the bath on the tracer diffusion coefficient of Fig. \ref{diff-tracer}. In this case, tracer diffusion is strongly coupled to the bath, and it becomes a collective phenomenon. 

\begin{figure}
\psfig{file=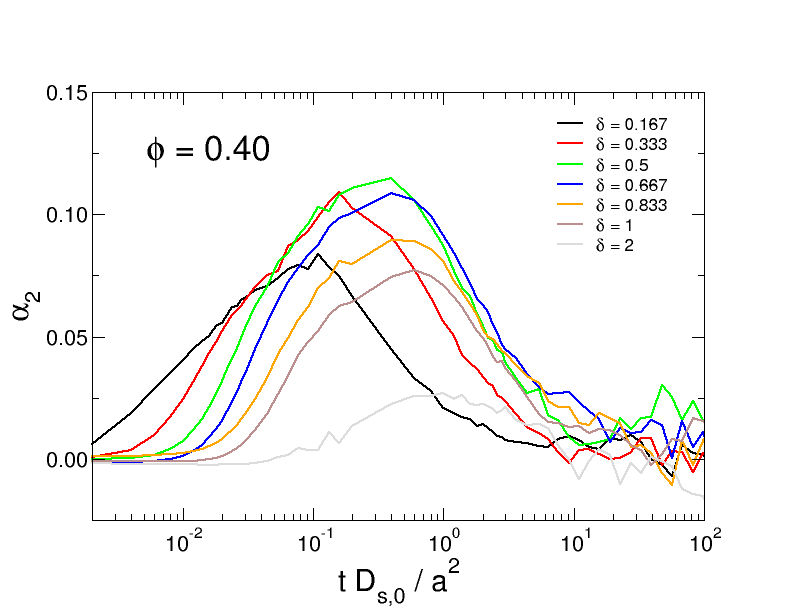,width=0.8\figurewidth}
\psfig{file=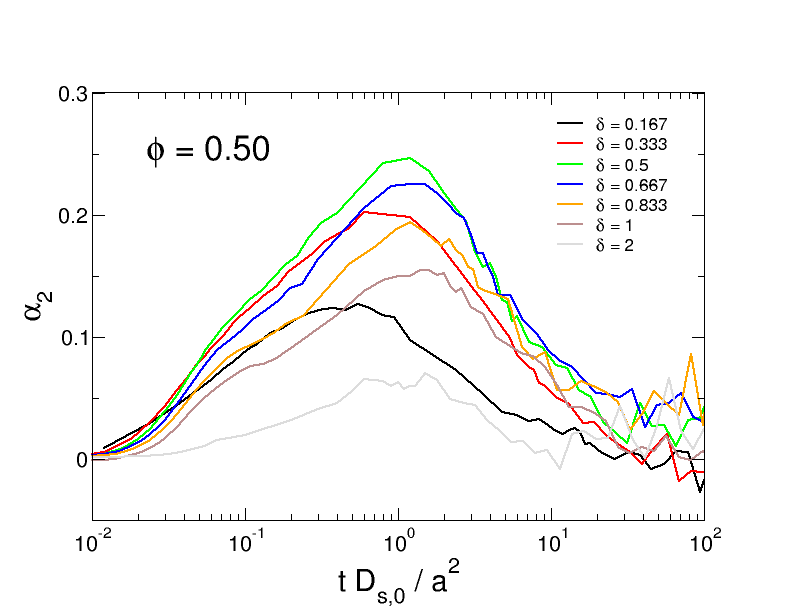,width=0.8\figurewidth}
\psfig{file=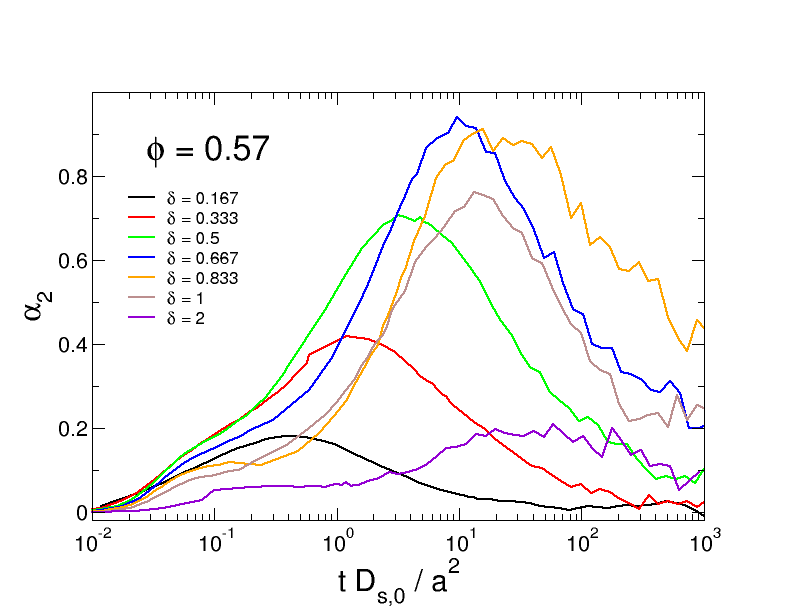,width=0.8\figurewidth}
\caption{Non-Gaussian parameter of the tracer particle, for different densities and tracer sizes (as labeled). \label{alpha2-tracer}}
\end{figure}

The non-Gaussian parameter calculated for the tracer displacement is shown in Fig. \ref{alpha2-tracer} for different tracer sizes; different panels correspond to different bath density. In all cases, $\alpha_{2,t}$ shows a maximum at intermediate times, but its height is larger than in the bulk (for the same density), for tracers slightly smaller than the bath particles; $\alpha_2$ for the bath is given by the case $\delta=1$. For tracers larger than the bath particles, $\delta > 1$, the non-Gaussian parameter is much smaller than the bulk value for all times, indicating that these tracers are not suitable for probing the heterogeneous dynamics of the bath.

\begin{figure}
\includegraphics[width=.9\figurewidth]{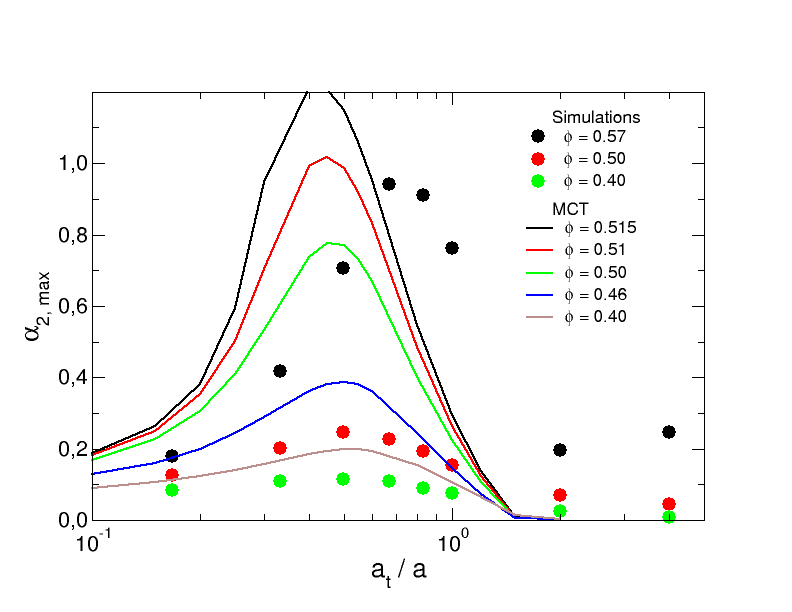}
\caption{Maximum of the tracer non-Gaussian parameter, for different densities (as labeled). \label{alpha2-max}}
\end{figure}

The maxima of $\alpha_{2,t}$ are plotted in Fig. \ref{alpha2-max} as a function of the tracer size, for different densities. As already discussed, this is maximal for tracers around half of the bath particles, increasing slightly with the bath density. Note that this maximum occurs for tracer sizes where the transition in the diffusion coefficient takes place (see Fig. \ref{diff-tracer}).

Qualitatively, \gls{mct} (lines) confirms the observation made in the simulation (points). There
is a size ratio, $\delta_*<1$, where the \gls{ngp} becomes maximal at any given density. Note however that the theory also predicts strongly negative minima preceding the maximum for $\delta>1$, a feature not observed in the simulation.

The size ratio of maximum heterogeneity in \gls{mct}, $\delta_*\approx0.5$,
is slightly smaller than the value suggested by the simulations,
$\delta_*\approx0.7$. This agrees with the qualitative deviation we have
observed for the time-dependent \gls{ngp} in general: the predictions of
the theory for a somewhat smaller $\delta$ match better what is observed
in the simulation.

\begin{figure}
\psfig{file=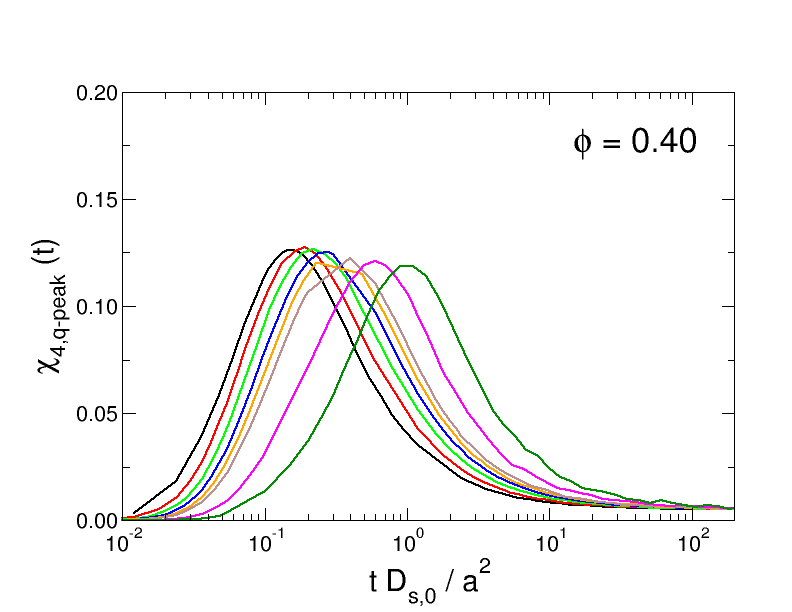,width=0.9\figurewidth}
\psfig{file=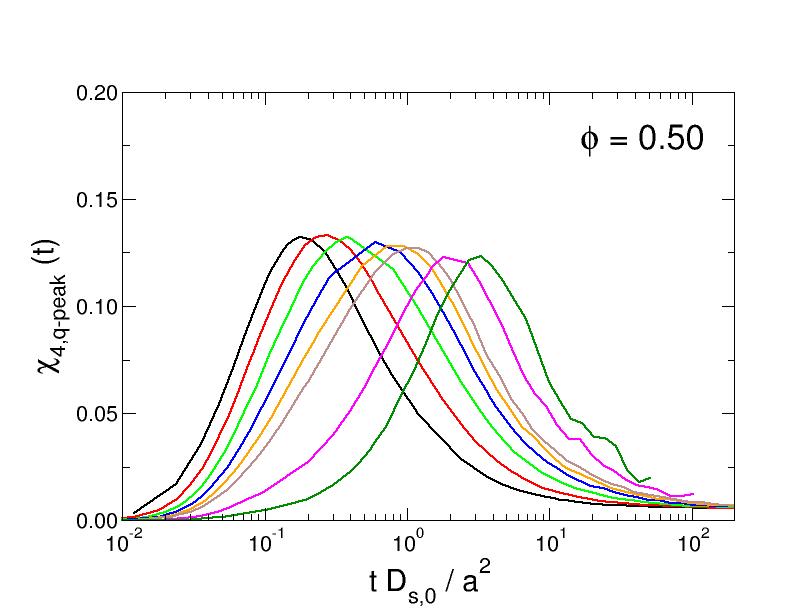,width=0.9\figurewidth}
\psfig{file=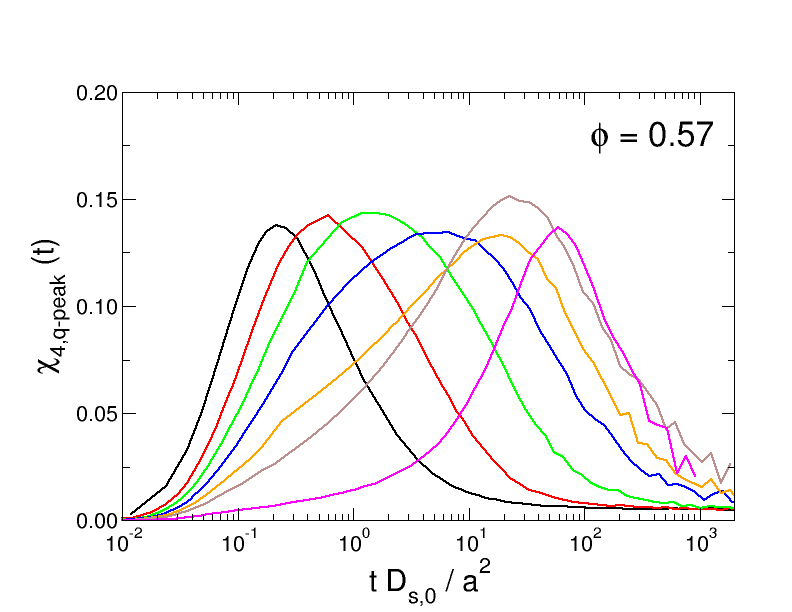,width=0.9\figurewidth}
\caption{Susceptibility of the tracer particle, for different densities and tracer sizes (as labeled). \label{chi4-tracer}}
\end{figure}

The susceptibility calculated from eq. (\ref{eq-chi4-tracer}) is shown in Fig. \ref{chi4-tracer} for the same cases as the non-Gaussian parameter. This parameter is much larger than its bulk counterpart (see Fig. \ref{chi4-bath}), because in that case an ensemble average is performed prior to the time-origin average; the former lacks in the tracer dynamics, resulting in a larger variability in the correlation function. The overall shape of $\chi_4^t$ is similar for all tracer sizes, with a maximum at intermediate times, and decaying to zero for short and long times. Also, although the maximum occurs at longer times for larger tracers, following the relaxation time of the intermediate scattering function, Fig. \ref{MSD-tracer}, its height is almost constant.

The comparison between $\alpha_2^t$ and $\chi_4^t$ shows that both have a maximum in the time corresponding to the relaxation time of the tracer scattering function, and both increase as the density is increased, and the glass transition is approached. However, the height of $\alpha_2^t$ is more sensitive to the bath dynamics, and bath-tracer coupling, making it a better parameter to determine the \gls{dh} with microrheology. Even more, the maximum value of $\alpha_2^t$ is obtained for tracers slightly smaller than the characteristic length scale of the bath.

\begin{figure}
\psfig{file=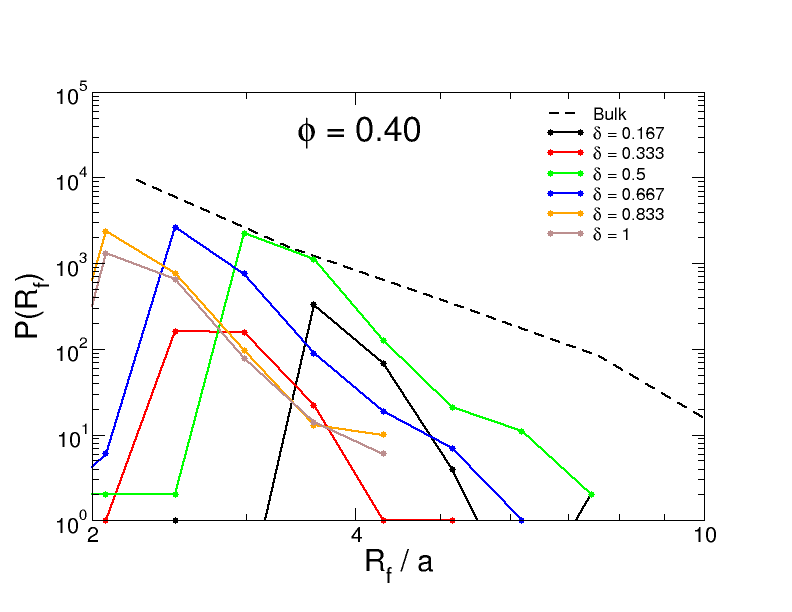,width=0.9\figurewidth}
\psfig{file=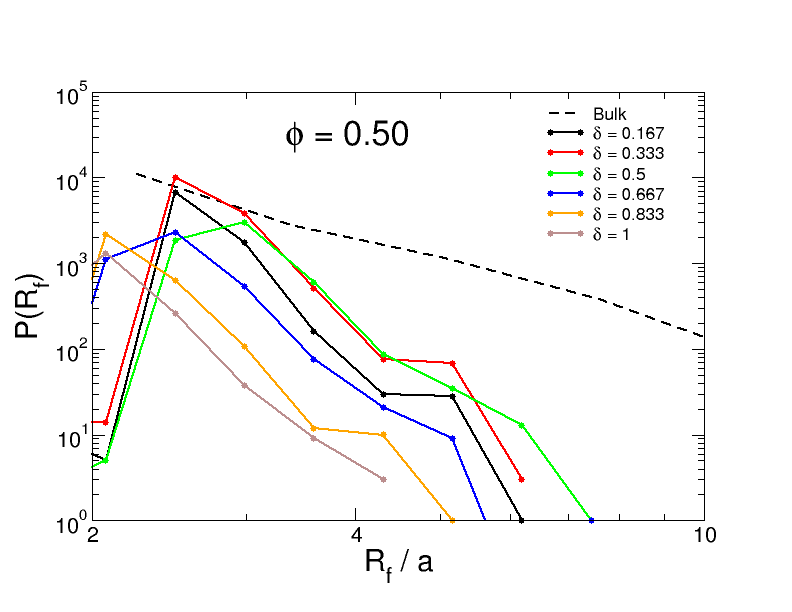,width=0.9\figurewidth}
\psfig{file=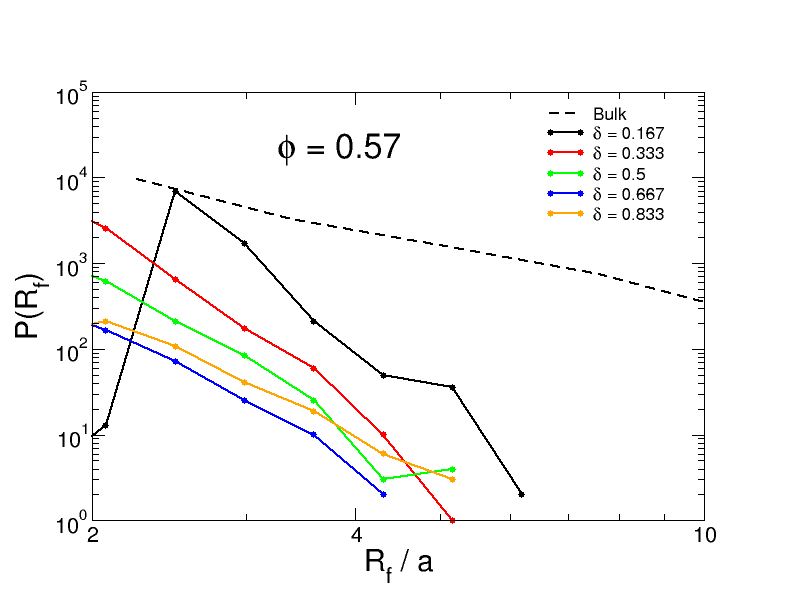,width=0.9\figurewidth}
\caption{Distribution of sizes of the fast regions determined with tracers of different sizes, as labeled. Different panels correspond to different bath densities. \label{fast-regions}}
\end{figure}


Finally, we test if the regions of different mobility can be directly probed with microrheology. As mentioned above, the squared displacement of the tracer particle in a time interval equal to the time of the maximal \gls{ngp} is calculated and this is compared with the mean value (tracer \gls{msd}). If it is larger than the MSD, it is assumed that the tracer is in a fast region, and its size, $R_f$, is determined as the distance from the entrance to the exit points. Obviously, this interpretation is appropriate only for small tracers, and this measurement underestimates the size of the fast regions as compared with the radius of gyration determined from the analysis of the bath particles dynamics (Fig. \ref{cluster-distrib-bulk}), because it is not guaranteed that the tracer crosses the whole region, but it serves to probe the existence of regions with different mobility. 

Fig. \ref{fast-regions} shows the distribution of distances $R_f$, for different tracer sizes and bath densities (different panels), in comparison with the cluster distribution of the bath. As expected, the latter extends to larger distances, but a dependence of the distribution with the tracer size can also be observed. Large tracer particles rarely belong to a cluster of fast particles, whereas small particles indeed show increased mobility over significant distances. Notably, the distribution observed for tracers of size equal to one half of the bath particles extends to larger distances for $\phi=0.40$, whereas smaller tracers must be considered for larger densities, in contrast to the heterogeneities in the tracer dynamics reported previously (see Fig. \ref{alpha2-max}). It is also interesting to note that the distributions have a peak at a finite distance of a few radii, that displaces to smaller values for larger tracers or bath density.

For tracers comparable to (but smaller than) the bath particles, the average distance traveled by the tracer during the period of increased mobility decreases upon increasing the bath density, contrary to the growth of the mean cluster size measured previously (see insets to Fig. \ref{cluster-distrib-bulk}) as well as the literature. Even for the smaller tracers, this trend is not corrected, indicating that the methodology is not valid to probe the size of the fast regions, but clearly indicate their presence in the systems under study.

\section{Conclusions}

In this work, we have proposed to use passive microrheology as a tool to grasp the \gls{dh} of model colloidal systems approaching the glass transition, using simulations and mode coupling theory. Tracers smaller and larger than the average bath particle have been used, and different parameters have been proposed, namely, the non-Gaussian parameter, the dynamic susceptibility, and the existence of regions of increased mobility. The different dynamics of small and large tracers has been observed, e.g. by the tracer self diffusion coefficient, transitioning from a regime where the tracer motion is decoupled from the bath dynamics for small tracers, to a strong coupling for large ones. 

The \gls{ngp} and dynamic susceptibility of the tracer mimic the overall behaviour of the bulk system, decaying to zero for small and long times, and increasing with the density. However, the dependence on the tracer size reveals that the \gls{ngp} is much more sensitive, and also allows to identify an optimal ratio of tracer size to bath particle size, where the effects are best observed (higher maximum in $\alpha_2$): the optimal size ratio is around $0.667$. These observations are confirmed by the theory semi-quantitatively, but the \gls{ngp} from the theory displays a notable negative dip at short times which is not reproduced in the simulations, presumably due to overestimating the confining effect of the cage of neighbors, which ultimately results in a quantitative error in the glass transition density. The dynamic susceptibility measured by microrheology, on the other hand, does not show a big dependence on the tracer size, but also on the bath density.

One further indication of the \gls{dh} is the presence of regions with different mobility, which has also been tested with microrheology. The method proposed here, based on identifying the parts of the tracer trajectory with faster-than-average dynamics, yields a decreasing size of the regions with increased mobility, contrary to the direct observations. Therefore, although passive microrheology shows the existence of these regions, and a clear dependence with the tracer size, this method cannot be regarded as providing reliable estimates of their dimension. 

Overall, the measurement of the \gls{ngp} is concluded as the more reliable and sensitive parameter to be measured with passive microrheology to probe the \gls{dh} in colloids close to the glass transition, with tracers slightly smaller than the bath particles. 

%
%
%
%

\section{Acknowledgements}

A.M.P. acknowledges financial support
from Grant No. PID2021-127836NBI00 funded by
MCIN/AEI/10.13039/501100011033/FEDER “A way to
make Europe”.

\bibliographystyle{apsrev}
\bibliography{references}

\end{document}